%% file: paper_oliva.tex
\documentclass{mem}
\usepackage{graphicx}
\usepackage[a4paper]{hyperref}
\usepackage{mdwlist} 

\begin{document}
\title
{The scientific use and productivity of the Telescopio Nazionale Galileo (TNG)}

  \author{E.~Oliva}

  \institute{
            INAF, Osservatorio Astrofisico di Arcetri, Largo E.~Fermi 5,
            I-51025, Firenze \email{oliva@arcetri.astro.it}
            \and
           INAF, Telescopio Nazionale Galileo, Apartado de Correos 565,
            E-38700 Santa Cruz de la Palma, Spain
            }

  \authorrunning{E.~Oliva}
  \titlerunning{Science with the TNG}

\abstract
{
This paper reviews the scientific use
of the TNG from the beginning of 
regular observations 
till the end
of 2007. Statistics are given for the time request, use and
productivity  of the telescope and its focal plane instruments.
Information on the down-times and a list of the major technical works/upgrades 
are also included.
}
\maketitle{}

\section{Introduction}
\label{introduction}
The Telescopio Nazionale Galileo (TNG) is a national facility
funded by the Italian Government through the Italian National Institute of
Astrophysics (INAF).
It consists of a $\oslash$3.58~m 
optical-infrared telescope with fixed, easily selectable focal plane 
instruments (see Sect.~\ref{instruments}).
It is located at the Roque de Los
Muchachos Observatory (La Palma, Spain) and has been operating for 
scientific observations since its first light in 1998. 
Scientific observing time is available to the whole international community
who can apply for observations to three different time allocation committees.
Following the international agreements and rules of the
La Palma Observatory, the observing time is distributed as follows.

%
      75\% of the time is selected and
      allocated by the INAF Time Allocation Committee (TNG-TAC). 
      Calls for proposals are issued twice a year, in March and September.
      The observations and scheduled are organized in two semesters, namely
      Feb-Jul (``$year$A") and Ago-Jan (``$year$B''). 
      For historical reasons, the observing campaigns  are also named 
      as ``AOT$i$'', with AOT1 starting at the beginning of 2000.
%
%

      20\% of the time is reserved to the Spanish community, through their
       ``Comisi\'on para la Asignaci\'on de Tiempo'' (CAT). The calls 
        and allocations are organized along the same semesters as the TAC
        time.
%
%

     5\%  is reserved to international projects (ITP) which make combined 
      use of
       various telescopes available at the Canary Island Observatories. 
       This time is allocated on annual basis by the  
      ``Comit\`e Cient\'\i fico Internacional'' (CCI)
%

\input{oliva_fig_tng}

\section{ TNG instrumentation }
\label{instruments}
The instruments available immediately after the first light were:\\
\begin{itemize*}
\item 
  OIG: an optical (U-Z bands) direct imager with a field of view of 
           4.9'$\times$4.9'. 
\item 
  ARNICA: a near infrared (J-K bands) imager with a f.o.v. of
    1.3'$\times$1.3'. It was decommissioned in Sept. 2000
\end{itemize*}

The actual, complete set of instruments was installed in the second
half of 2000.  
It consists of
\begin{itemize*}

\item DOLORES: an optical (U-Z bands) multi-mode, single-channel
instrument with a field
of view of 8.6'$\times$8.6' and long-slit spectroscopic 
capabilities with
resolving power (for 1" slit) ranging from 500 to 7000. It also
includes a multi-slit mode using masks produced with a punching machine.

\item SARG: a high resolution ($R_{\max}=164,000$)
cross-dispersed spectrometer
covering the 4000-9000 \AA\  wavelength range and especially designed
for high accuracy radial velocity measurements. It also includes a 
polarimetric module.

\item NICS: a near infrared (Y--K bands) multi-mode instrument with a field
of view of 4.3'$\times$4.3' and long-slit spectroscopic  capabilities
with resolving power (for 1" slit) ranging from 40 to 1500.  
Imaging-polarimetry and spectro-polarimetry modes are also available.
\end{itemize*}
All the instruments are normally mounted on the two Nasmyth foci 
(see Fig.~\ref{tng_mosaic}) 
and can
be rapidly selected by moving a few flat mirrors. Instruments are removed
only in case of maintenance operations which cannot be performed otherwise.
The largest instruments (DOLORES and SARG) have never been removed
since their first installation. 

\input{oliva_fig_queuing}

The telescope also includes an adaptive optics module (AdOpt) in a 
dedicated interface which takes all the space, weigh and momentum
available at the first Nasmyth derotator.
When inserted in the optical path (by moving a few flat mirrors)
it feeds NICS yielding
images with a 3$\times$ expanded scale, suitable to properly sample the
diffraction limit of the telescope.
The AdOpt module also included a Speckle camera operating at optical
wavelengths, this instrument was decommissioned in 2003.  \\

\section{Scheduling and service mode}
\label{scheduling}

Thanks to the wide range of observing mode offered by the focal plane 
instruments, their apparently rigid setup effectively translates into a 
remarkable flexibility of the scientific operations. In particular, it 
allows queuing scheduling of programs, where the observations are
prioritized  and
executed when the system technical status and meteorological conditions  
are appropriate. 

Starting from February 2003, TNG has regularly performed service 
observations and queuing scheduling for proposals approved by the 
Italian Time Allocation Committee (TAC). 
The proposals scheduled in queuing
are divided and prioritized in three categories (A, B, C) depending on
their scientific merit. 

A fourth, top priority category (S) was experimentally added between 
2005 and 2007
to favour the science verification programs for the adaptive
optics endorsed by INAF. 
All the proposals for AdOpt approved by the TNG-TAC were therefore
queued with higher priority than the remaining programs, and executed
on a best effort basis.

Fig.~\ref{fig_queuing} summarizes the efficiency of queuing observations.
The fraction of completion of class-A programs is always above 80\%,
and in most cases above 90\%. The average figures for class-B (71\%)
and class-C (44\%) programs are also very good. Due to the much 
lower efficiency of 
class-S programs it has been decided to separate the
scheduling of AdOpt science verification from normal 
service/queuing observations.
 
The fraction of time scheduled
in service and/or queuing mode has been about 45\% of the total
time allocated by the TAC.

\input{oliva_fig_downtime_aot}

\section{Technical works}
\label{technical}

The major technical operations performed after 
the TNG commissioning phase (\cite{bortoletto2001}) were:

\begin{itemize*}
\item Installation and first light of DOLORES, SARG and NICS (2000) 
\item Refurbishment of the DOLORES mechanics (2001) 
\item Re-alignment and activation of the compensation correction of the 
  telescope encoders (2002-2003) 
\item Substitution of the support bearings of the rotating building (2003) 
\item Refurbishment of the NICS mechanics (2003) 
\item Refurbishment of the AdOpt module (2004-2005) 
\item Upgrade of the derotators optics control system (2004-2006) 
\item Upgrade of the Telescope Active Optics control system (2004-present) 
\item Refurbishment of the NICS array control system (2006) 
\item Upgrade of the auto-guide system (2006-2007) 
\item Upgrade of the DOLORES detector, interface and instrument control system 
   (2006-present) 
\item Refurbishment of the tacho-generator of the Telescope motors 
      (2006-present) 
\item Substitution and re-phasing of damaged telescope motors (2006-present)
\end{itemize*}
\input{oliva_fig_weather}

In practice, TNG underwent a general technical upgrade comparable to
the NTT big-bang (\cite{ntt_big_bang}) maintaining, however, full
operativity for science observations.
This remarkable result was possible thanks to a quite large fraction of
technical time ($\simeq$20\% of the total) which was effectively
coupled with the service observations, 
always guaranteeing that any night-time which remained 
available was used for scientific observations.

\section{Downtimes}
\label{downtimes}

The downtimes averaged over the observing campaigns/semesters
are summarized in Fig.~\ref{fig_downtime_aot}. 
The technical downtime is defined as the time lost for technical problems
which could not be re-scheduled. The decrease in 2003 coincides with 
the beginning of the service observations, which allowed a more flexible
organization of the technical time and (re-)scheduling of the programs.
 
The somewhat large fraction of time lost for meteorological reasons is not
evenly distributed through the year.
While the summer is extremely good (fraction of time lost $\simeq$7\%), 
however, downtimes of 45\% are typical between November and February
(see Fig.~\ref{weather}). 
%
%
\input{oliva_fig_proposals}

The seasonal behaviour of weather downtimes affected in very
different ways observing programs targeting galactic 
(visible in summer) 
and extra-galactic objects. 
Although the service mode with flexible 
scheduling made it possible to complete some of the programs targeting objects
visible in winter, however, the general effect on the astronomical 
community was a decrease of proposals for wintry objects and an
increase of the fraction of those specifically requesting the summer.

\section{Observing proposals}
\label{proposals}
\input{oliva_fig_publications}

Fig.~\ref{fig_proposals} shows the statistics and time evolution of the
observing proposals submitted to the TNG-TAC. The
proposals have decreased by about a factor of two between 2001 and 2006,
and then flattened to a roughly constant value of 400 requested nights per
year. 
The
over-subscription factor has varied from 3.8 to 1.8, but still remains one
of the largest among telescopes of similar class. 
 
The most requested instrument is DOLORES,  
which acts as the workhorse for all the programs requiring
imaging and/or low-medium resolution spectroscopy at visual wavelengths.
The average number of nights requested by each program has remained 
$\simeq$2.2 up until 2006, but increased to almost 3 in the following
years. This positive trend reflects the increasing importance of large
programs. 

The time requests for SARG have remained roughly constant, and even
 increased in the last years, approaching the values of DOLORES. 
The SARG programs are fewer but typically request a larger number of 
nights than the other instruments.
The average number of nights per
proposal remained $\simeq$3.5 between 2000 and 2007. The jump to 6.3
in 2008 coincides with the submission of a proposal for 
a large program which requires 60 nights/year.
 
NICS is the second most requested instrument in terms of number of proposals.
However, the time requests after 2005 have been lower than those for SARG.
The average time requested has always remained below 2.5 nights per program.
 
The time requests for OIG and AdOpt are much lower than for the other 
instruments.  The OIG graph does not extend to 2008 because the instrument has
been decommissioned.\\


Starting from 2003, a policy for long term programs
was encouraged by INAF 
and implemented by the TNG Time Allocation Committee. The fraction of time
allocated to long term programs has remained between 10 and 20\%.

\section{Scientific publications}

The first, complete list of the refereed papers based on data collected 
at the TNG was compiled in 2005 (\cite{boschin2005}). This list is 
periodically updated 
and the statistics of publications are summarized in 
Fig.~\ref{fig_publications}
which also includes the citations (from the ADS database) and the
distribution of publications per instruments.

The first and most encouraging result is the constant increase of the number
of publication which in 2007 surpassed the remarkable figure of 1 publication
per week. Particularly positive is also the comparison between the total
numbers of publications and of observing nights since the beginning of TNG
scientific operation, which yields an efficiency rate of about
0.8 publications per week of scheduled time.

The evolution of the citations shown in upper-right panel of 
Fig.~\ref{fig_publications} is characterized by a peak of almost 730 citations
for the papers published in 2003. The values for the subsequent years are
lower by a factor of about two. This decrease can be (at least partly)
attributed to the cumulative effect of the citation index, which tends 
to favour older publications.
The average number of citations per publication is 15. The efficiency
rate is close to 1.6 citations per night of scheduled time.

The distribution of publications and citations among the various instruments
(lower panels of Fig.~\ref{fig_publications}) substantially reflects their use.
In other words, the ratio between number of publications/citations 
and scheduled time is roughly the same among the instruments. The only
exception is the adaptive optics module which was scheduled for about
5\% of the time  but did not produce refereed publications.

Listed below are the papers with the largest number ($\ge$60) of
citations (from the ADS database, updated on May 2008)\\

%
%
\cite{santos2004} 
  {\it Spectroscopic [Fe/H] for 98 extra-solar 
  planet-host stars.  Exploring the probability of planet formation }
  Based on TNG-SARG and ESO-FEROS spectra,
  206 citations. Similar works by Santos et al. 
  (\cite{santos2003},\cite{santos2005}) have 
  138 and 69 citations, respectively.

\cite{fiore2003}
    {\it The HELLAS2XMM survey. IV. Optical identifications and the 
    evolution of the accretion luminosity in the Universe}.
    Based on deep images taken with DOLORES-TNG and EFOSC2-ESO,
    131 citations.

\cite{fymbo2001}
    {\it Detection of the optical afterglow of GRB 000630: Implications 
    for dark bursts}.
    Based on OIG-TNG images in combination with data from
    the Calar-Alto, USNO, and NOT telescopes; 96 citations.

\cite{gratton2003}
    {\it Abundances for metal-poor stars with accurate parallaxes. 
    I. Basic data}. Based on spectra taken with SARG-TNG, UVES-VLT and 
    with the McDonald 2.7m telescope, 85 citations

\cite{natta2002}
   {\it Exploring brown dwarf disks in rho Ophiuchi}.
   Based on spectra taken with NICS-TNG, 68 citations

\cite{tagliaferri2005}
   {\it GRB 050904 at redshift 6.3: observations of the oldest cosmic 
   explosion after the Big Bang}.
   Based on NICS-TNG images in combination with 
   ESO and Calar-Alto imagers, 64 citations

\cite{maiolino2004}
   {\it A supernova origin for dust in a high-redshift quasar}.
   Based on NICS-TNG spectra,  62 citations

\cite{masetti2000}
   {\it Unusually rapid variability of the GRB000301C optical afterglow}
    Based on OIG-TNG images in combination with data from
    Calar-Alto, Loiano, UPSO and SNO telescopes; 62 citations.

\input{oliva_biblio}

\end{document}

%% file: oliva_fig_tng.tex
\begin{figure}[t]
\centerline{
\includegraphics[width=\hsize]{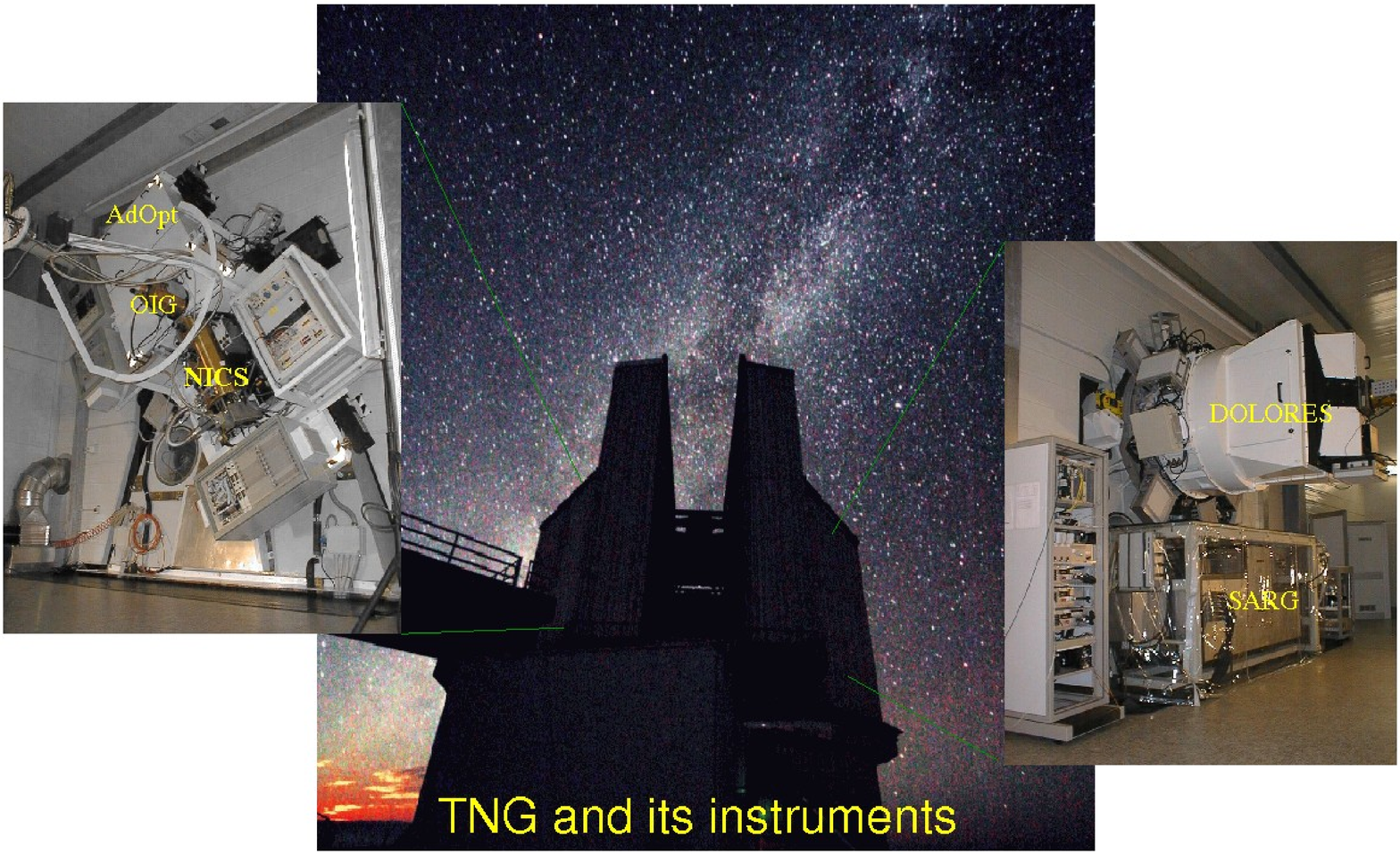}
}
\label{tng_mosaic}
\caption{
The TNG and its instruments.
}
\end{figure}

%% file: oliva_fig_queuing.tex
\begin{figure}[t]
\centerline{
\includegraphics[width=\hsize]{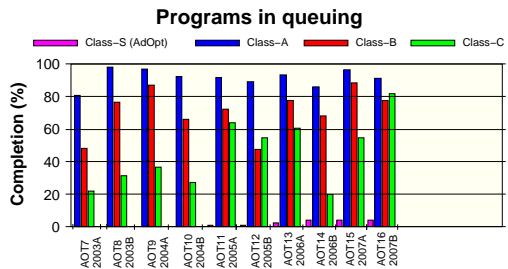}
}
\label{fig_queuing}
\caption{
Completion statistics of TNG programs scheduled in queuing mode in the
semesters since the beginning of this observing mode in 2003.
}
\end{figure}

%% file: oliva_fig_downtime_aot.tex
\begin{figure}[t]
\label{fig_downtime_aot}
\centerline{
\includegraphics[width=\hsize]{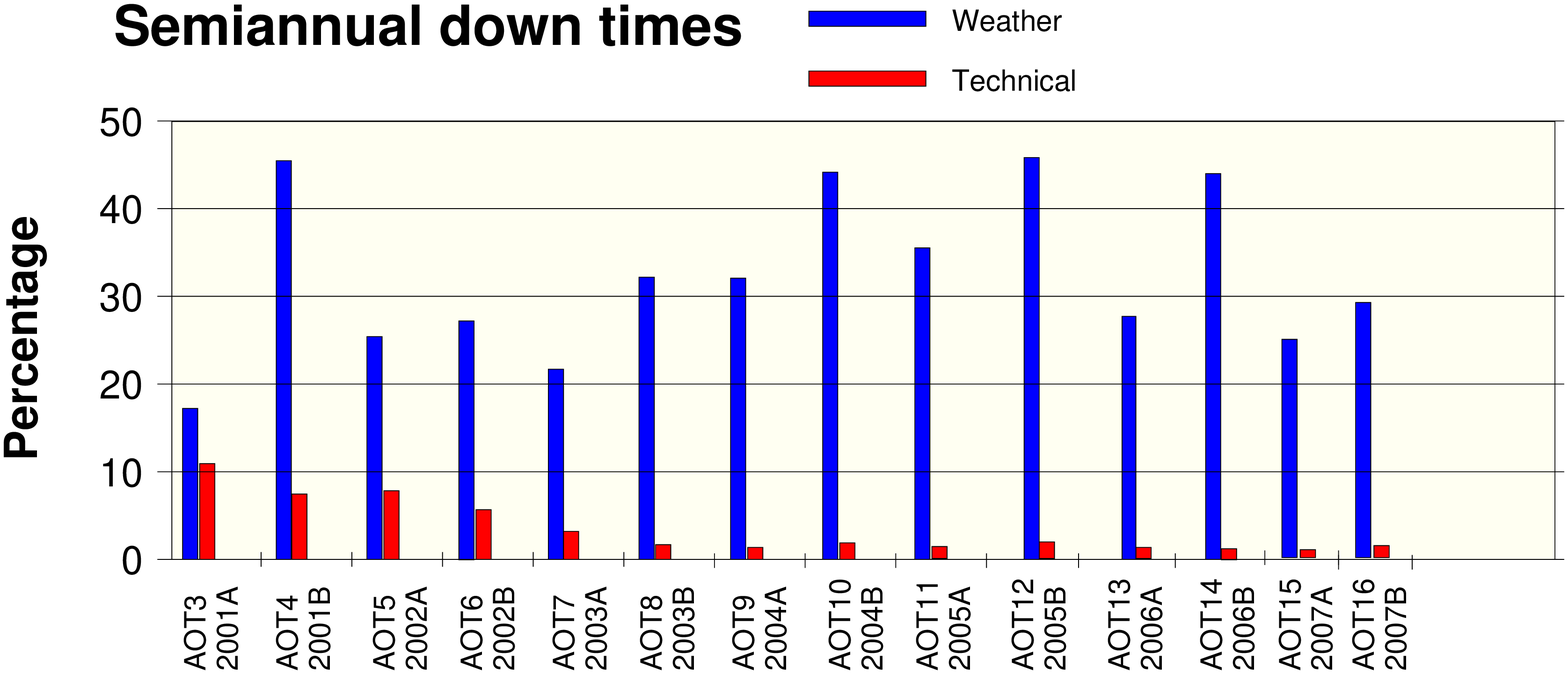}
}
\caption{
Downtime statistics for TNG. The values are averaged over the Feb-Jul ("A") and
Ago-Jan ("B") observing semesters/campaigns.
}
\end{figure}

%% file: oliva_fig_weather.tex
\begin{figure}[t]
\centerline{
\includegraphics[width=\hsize,height=0.25\vsize]{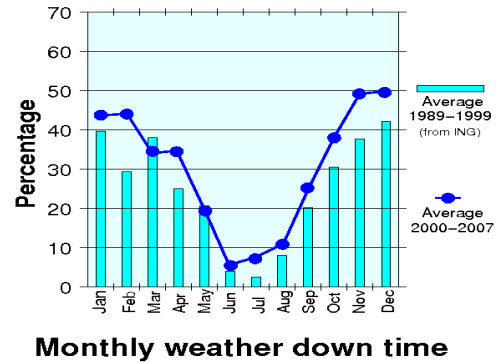}
}
\label{weather}
\caption{
Fraction of observing time lost for meteorological reasons. 
}
\end{figure}

%% file: oliva_fig_proposals.tex
\begin{figure}[t]
\centerline{
\includegraphics[width=\hsize]{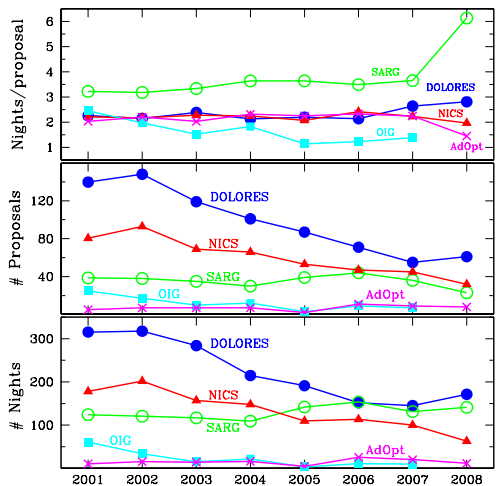}
}
\label{fig_proposals}
\caption{
Evolution of proposals for observations submitted to the TNG-TAC. 
}
\end{figure}

%% file: oliva_fig_publications.tex
\begin{figure*}[ht]
\centerline{
\includegraphics[width=\hsize]{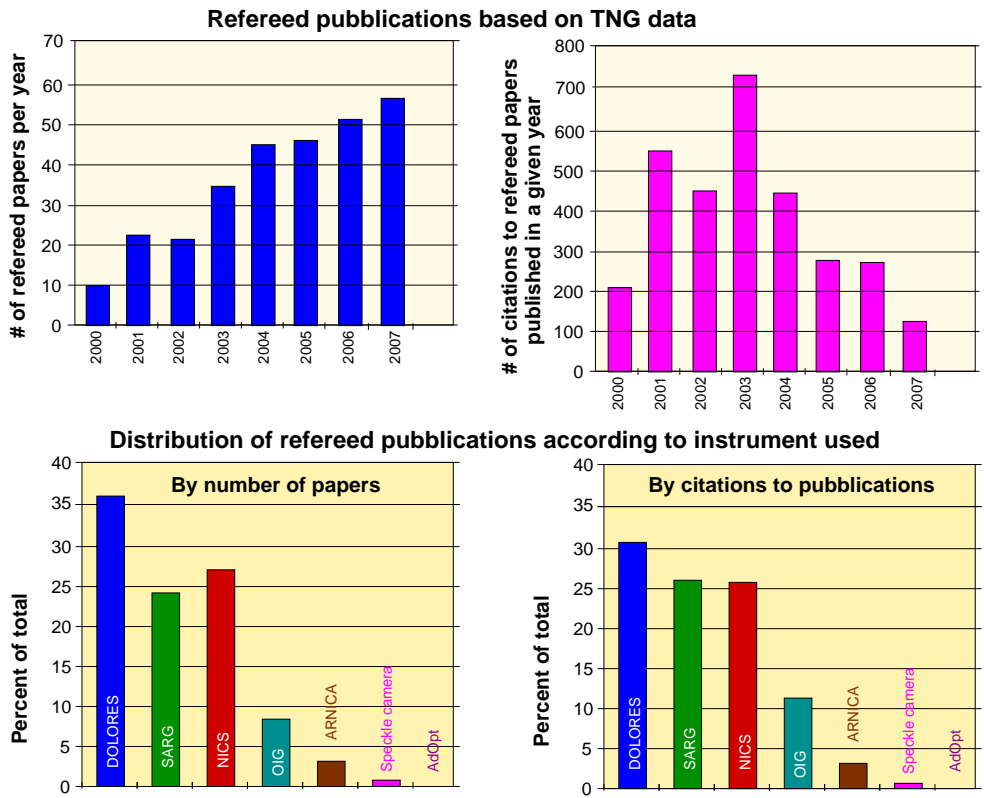}
}
\label{fig_publications}
\caption{
Statistics of publications based on data collected at TNG.
The numbers include papers published in refereed journals up to the end
of 2007.
}
\end{figure*}

%% file: oliva_biblio.tex
\bibliographystyle{aa}

%% file: paper_oliva.bbl
\begin{thebibliography}{}

\bibitem[Bortoletto, 2001]{bortoletto2001}
Bortoletto, F.; 2001, New Astronomy Rev., 45,37

\bibitem[Boschin, 2005]{boschin2005}
Boschin, W.; 2005, astro-ph/0604271

\bibitem[Fiore et al. (2003)]{fiore2003} 
Fiore, F.; Brusa, M.; Cocchia, F.; Baldi, A.; Carangelo, N.; Ciliegi, P.;
Comastri, A.; La Franca, F.; Maiolino, R.; Matt, G.; et al.; 
2003, A\&A, 409, 79

\bibitem[Fymbo et al. (2001)]{fymbo2001} 
Fynbo, J. U.; Jensen, B. L.; Gorosabel, J.; Hjorth, J.; Pedersen, H.; 
Møller, P.; Abbott, T.; Castro-Tirado, A. J.; Delgado, D.; Greiner, J.; et al.;
2001, A\&A, 369, 373

\bibitem[Gratton et al. (2003)]{gratton2003} 
Gratton, R. G.; Carretta, E.; Claudi, R.; Lucatello, S.; Barbieri, M.;
2003,  A\&A, 404, 187

\bibitem[Maiolino et al. (2004)]{maiolino2004} 
Maiolino, R.; Schneider, R.; Oliva, E.; Bianchi, S.; Ferrara, A.; 
Mannucci, F.; Pedani, M.; Roca Sogorb, M.;
2004, Nature, 431, 553

\bibitem[Masetti et al. (2000)]{masetti2000} 
Masetti, N.; Bartolini, C.; Bernabei, S.; Guarnieri, A.; Palazzi, E.; 
Pian, E.; Piccioni, A.; Castro-Tirado, A. J.; Castro Cer\'on, J. M.; 
Verdes-Montenegro, L.; et al.;
2000, A\&A, 359, 23

\bibitem[Natta et al. (2002)]{natta2002} 
Natta, A.; Testi, L.; Comerón, F.; Oliva, E.; D'Antona, F.; Baffa, C.; 
Comoretto, G.; Gennari, S.; 
2002,
A\&A, 393, 597

\bibitem[2003]{santos2003} 
Santos, N.C.; Israelian, G.; Mayor, M.; Rebolo, R.; Udry, S.; 2003,
A\&A, 398, 363

\bibitem[Santos et al. (2004)]{santos2004} 
Santos, N.C.; Israelian, G.; Mayor, M.; 2004,
A\&A, 415, 1153

\bibitem[2005]{santos2005} 
Santos, N. C.; Israelian, G.; Mayor, M.; Bento, J. P.; Almeida, P. C.; 
Sousa, S. G.; Ecuvillon, A.;
2005, A\&A, 437, 1127

\bibitem[Spyromilio, 1997]{ntt_big_bang}
Spyromilio, J.; 1997, The Messenger, 88, 12

\bibitem[Tagliaferri et al. (2005)]{tagliaferri2005} 
Tagliaferri, G.; Antonelli, L. A.; Chincarini, G.; 
Fern\'andez-Soto, A.; 
Malesani, D.; Della Valle, M.; D'Avanzo, P.; Grazian, A.; Testa, V.; 
Campana, S.; et al.;
2005, A\&A, 443, L1


\end{thebibliography}
